\documentclass{aa} 

\DeclareRobustCommand{\rchi}{{\mathpalette\irchi\relax}}
\newcommand{\irchi}[2]{\raisebox{\depth}{$#1\chi$}} 

\newcommand{\ex}{\mathbf{e}_{\rm x}}
\newcommand{\ey}{\mathbf{e}_{\rm y}}
\newcommand{\ez}{\mathbf{e}_{\rm z}}

\newcommand{\rlight}{r_{\rm L}}

\newcommand{\ephi}{\mathbf{e}_\varphi}

\DeclareMathOperator{\arccot}{arccot}

\usepackage{graphicx}
\usepackage{txfonts}
\usepackage{color}

\begin{document} 

\title{Spheroidal magnetic stars rotating in vacuum}


\author{J. P\'etri
}

\institute{Universit\'e de Strasbourg, CNRS, Observatoire astronomique de Strasbourg, UMR 7550, F-67000 Strasbourg, France.\\
\email{jerome.petri@astro.unistra.fr}         
}

   \date{Received ; accepted }

 
  \abstract
   {Gravity shapes stars to become almost spherical because of the isotropic nature of gravitational attraction in Newton's theory. However, several mechanisms break this isotropy like for instance their rotation generating a centrifugal force, magnetic pressure or anisotropic equations of state. The stellar surface therefore deviates slightly or significantly from a sphere depending on the strength of these anisotropic perturbations.}
   {In this paper, we compute analytical and numerical solutions of the electromagnetic field produced by a rotating spheroidal star of oblate or prolate nature. This study is particularly relevant for millisecond pulsars for which strong deformations are produced by the rotation or a strong magnetic field, leading to indirect observational signatures of the polar cap thermal X-ray emission.}
   {First we solve the time harmonic Maxwell equations in vacuum by using oblate and prolate spheroidal coordinates adapted to the stellar boundary conditions. The solutions are expanded in series of radial and angular spheroidal wave functions. Particular emphasize is put on the magnetic dipole radiation. Second, we compute approximate solutions by integrating numerically the time-dependent Maxwell equations in spheroidal coordinates.}
   {We show that the spin down luminosity corrections compared to a perfect sphere are to leading order given by terms involving $(a/\rlight)^2$ and $(a/R)^2$ where $a$ is the stellar oblateness or prolateness, $R$ the smallest star radius and $\rlight$ the light-cylinder radius. The corresponding perturbations in the electromagnetic field are only perceptible close to the surface, deforming the polar cap rims. At large distances $r\gg a$, the solution tends asymptotically to the perfect spherical case of a rotating dipole.}
   {}


\maketitle

%

\section{Introduction}

Large celestial bodies are approximately spherical because of the preponderance of gravity against other internal forces and stresses. Gravitation does not favour any direction in the sky and therefore a perfect spherical shape is expected for isotropic materials. However, celestial bodies like stars and molecular clouds are often subject to rotation, producing a centrifugal force $F_{\rm cen}$ breaking the isotropy imposed by gravity~$F_{\rm grav}$. A reference direction appears along the rotation axis. The surface of the body is deformed and can be approximated for instance by an ellipsoidal shape of oblate nature. The strength of this force must be compared to gravity at the surface. A good guess for this ratio is given by the comparison between centrifugal and gravitational forces
\begin{equation}\label{eq:Fcentrifuge}
\frac{F_{\rm cen}}{F_{\rm grav}} \approx \frac{\Omega^2}{\Omega_k^2}
\end{equation}
$\Omega$ being the rotation rate of the star (assumed to be in solid body rotation) and $\Omega_k$ the Keplerian angular frequency at the equator. For a more quantitative description, see \cite{chandrasekhar_ellipsoidal_1970} discussion about spheroids and ellipsoids of revolution and \cite{horedt_polytropes:_2004} who presents a comprehensive analysis of rotational effects on polytropic stars. Centrifugal forces are particularly important during the violent birth of a neutron star, where they deform the proto-neutron star to an oblate shape. The study proposed in this paper will be relevant to such early infancy of a newly born neutron star.

For fast rotating stars, with rotation rate being a fraction of the equatorial Keplerian frequency, we expect its shape to deviate from a spherical body by a fraction given by eq.\eqref{eq:Fcentrifuge}.
Following \cite{zanazzi_electromagnetic_2015}, the rotation-induced oblateness of a celestial corps of uniform density is estimated analytically through the parameter
\begin{equation}\label{eq:epsilon}
\epsilon = \frac{15}{16\, \pi} \, \frac{\Omega^2}{G \, \rho} = \frac{5}{4} \, \frac{\Omega^2 \, R^3}{G \, M} = \frac{5}{4} \, \frac{\Omega^2}{\Omega_k^2}
\end{equation}
defined by the moment of inertia difference between equatorial and polar direction. This is in agreement with the estimate derived in eq.\eqref{eq:Fcentrifuge}. It gives an estimate (likely an overestimate) for the oblateness ratio $a/R$. Exact analytical solutions exist for a constant density and uniformly rotating axisymmetric ellipsoidal fluid. If the ellipticity is defined as
\begin{equation}\label{eq:Ellipticite}
e = \sqrt{1 - \left(\frac{R_{\rm pol}}{R_{\rm eq}}\right)^2}
\end{equation}
with $R_{\rm pol}$ and $R_{\rm eq}$ being the polar and equatorial radii of the fluid, frequency~$\Omega$ given by \cite{chandrasekhar_ellipsoidal_1970} reads
\begin{equation}\label{eq:MacLaurin}
\frac{\Omega^2}{2\,\pi\,G\,\rho} = \frac{\sqrt{1-e^2}}{e^3} \, (3-2\,e^2) \, \arcsin e - 3 \,  \frac{1-e^2}{e^2} \approx \frac{4}{15} \, e^2 .
\end{equation}
The approximation is valid for small ellipticities $e\ll1$. A more realistic estimate is obtained for instance for the SLy4 equation of state as found from Table~1 of \cite{silva_surface_2021}.

The body becomes oblate and impacts its immediate surrounding gravitationally but also electromagnetically if it possesses a magnetic field. We are interested in the latter possibility namely an oblate magnetic star rotating in vacuum. It launches an electromagnetic wave responsible for its magnetic braking. Strongly magnetized neutron stars are of particular interest with this respect because they spin down according to the magnetodipole losses. Millisecond pulsars are the fastest rotating neutron stars known, they become elongated along their equator because of strong centrifugal forces. It is therefore important to understand how their oblate shape perturbs the electromagnetic field and Poynting flux in conjunction with their ageing and spin evolution. So far exact analytical solutions only exist for a perfect sphere as expressed by \cite{deutsch_electromagnetic_1955}. Extension to multipolar fields have been thoroughly investigated by \cite{petri_multipolar_2015}, see also \cite{bonazzola_general_2015}. It is our goal to extend these important results to oblate stars by introducing a coordinate system adapted to the stellar surface shape. Consequently oblate spheroidal coordinates are best suited to achieve this goal. These coordinates are among the eleven separable systems \citep{morse_methods_1953-1}, leading to well defined solutions for the Laplace and Helmholtz equations. For completeness, we will also consider prolate shapes although these cannot be produced by rotation, but for instance by magnetic pressure.
Indeed, observational signatures of such prolateness is witnessed by the torque-free precession of magnetars subject to strong toroidal magnetic fields \citep{makishima_evidence_2016, makishima_nustar_2019}.

For strongly magnetized systems, the magnetic pressure becomes comparable to the gaseous pressure and deforms its surface to an aspherical shape with no particular symmetry axis. Spheroidal geometries could therefore be used as a first step towards more general configurations. Spheroidal coordinate systems possess the interesting and important properties of fully separation of variable for the Laplace and Helmholtz operators. It is therefore possible to solve analytically time harmonic wave emission and propagation in spheroidal coordinates. Moreover for compact objects, anisotropic equation of states for nuclear matter at very high density like in neutron stars also produces non spherical astronomical objects. 

From a mathematical perspective, spheroidal wave functions applied to electromagnetic theory have been extensively discussed in \cite{li_spheroidal_2001}. Some early application to light scattering was studied by \cite{asano_light_1975}. A different approach employing only spheroidal coordinates has been proposed by \cite{zeppenfeld_solutions_2009}.


In this paper, we compute formally exact analytical solutions to the electromagnetic wave radiation of a stationary rotating star of spheroidal shape, oblate or prolate. In section~\ref{sec:spheroidal}, we recall the useful and important properties of the curvilinear and orthogonal coordinate systems formed by oblate and prolate coordinates, with their metric and natural basis. The separation of variables is presented and the related spheroidal wave functions are introduced. Next in section~\ref{sec:static}, we compute static solutions for the multipolar magnetic field sustained by a spheroidal object. We will discuss the important question about the normalization of a spheroidal multipole with respect to a spherical multipole taking as the reference solution. Eventually, in section~\ref{sec:rotating} we compute the solution for a rotating spheroid by expansion of the electromagnetic field into spheroidal wave functions. Some useful approximate analytical solutions are presented to the lowest order in oblateness or prolateness. We close our work with accurate numerical results of spheroidal stars rotating in vacuum performed by pseudo-spectral time-dependent simulations in section~\ref{sec:simulations}. Conclusions are drawn in section~\ref{sec:conclusion}.


\section{Spheroidal coordinates}
\label{sec:spheroidal}

We start with a brief overview of both spheroidal coordinate systems, namely oblate and prolate.
Let us assume a star with minimal radius~$R$, corresponding to the polar radius for oblate coordinates and to the equatorial radius for prolate coordinates. An oblate shape describes well a self-gravitating gas deformed by its own rotation. For completeness, we also consider prolate shapes although these deformations are not produced by rotation. It is advantageous to adapt the curvilinear coordinate system to the boundary conditions imposed by the gas or the star.

The Cartesian coordinate system is defined by the unit orthonormal basis $(\ex, \ey, \ez)$ with the associated coordinates $(x,y,z)$. We define the spheroidal coordinates relying on the Cartesian correspondence.

\subsection{Oblate spheroidal coordinates}

We introduce the oblate spheroidal coordinate system $(\rho,\psi,\varphi)$ such that the Cartesian coordinates $(x,y,z)$ are given by
\begin{subequations}
	\label{eq:oblate}
	\begin{align}
	x & = \eta \, \sin \psi \, \cos\varphi \\
	y & = \eta \, \sin \psi \, \sin\varphi \\
	z & = \rho \, \cos \psi
	\end{align}
\end{subequations}
where we define $\eta=\sqrt{\rho^2+a^2}$ and $a$ is a real and positive parameter related to the oblateness. The equatorial radius of the surface is then $R_{\rm eq} = \sqrt{R^2+a^2}$.
The ellipticity is defined by \citep{shapiro_black_1983}
\begin{equation}\label{eq:ellipticityOblate}
\epsilon = \frac{R_{\rm eq} - R}{(R_{\rm eq} + R)/2} .
\end{equation}
The natural basis vectors derived from eq.\eqref{eq:oblate} are expressed as
\begin{subequations}
	\label{eq:Base_naturelle}
	\begin{align}
	\eta \, \vec{e}_\rho & = \rho \, \sin \psi \, \cos \varphi \, \ex + \rho \, \sin \psi \, \sin \varphi \, \ey + \eta \, \cos \psi \, \ez \\
	\vec{e}_\psi & = \eta \, \cos \psi \, \cos \varphi \, \ex + \eta \, \cos \psi \, \sin \varphi \, \ey - \rho \, \sin \psi \, \ez \\
	\vec{e}_\varphi & = - \eta \, \sin \psi \, \sin \varphi \, \ex + \eta \, \sin \psi \, \cos \varphi \, \ey .
	\end{align}
\end{subequations}
The position vector~$\vec{r}$ is then expanded into
\begin{equation}
\vec{r} = \frac{\rho\,\eta^2}{\Delta_o} \, \vec{e}_\rho + \frac{a^2\,\cos\psi\,\sin\psi}{\Delta_o} \, \vec{e}_\psi
\end{equation}
with $\Delta_o = \rho^2 + a^2 \, \cos^2 \psi$.
The oblate coordinate system is orthogonal, thus leading to a diagonal metric~$g$ with coefficients
\begin{subequations}
	\begin{align}
	g_{\rho\rho} & = \frac{\rho^2 + a^2 \, \cos^2 \psi}{\eta^2} = \frac{\Delta_o}{\eta^2}\\
	g_{\psi\psi} & = \rho^2 + a^2 \, \cos^2 \psi = \Delta_o \\
	g_{\varphi\varphi} & = \eta^2 \, \sin^2 \psi .
	\end{align}
\end{subequations}
Its determinant is
\begin{equation}
\gamma = \det(g) = (\rho^2 + a^2 \, \cos^2 \psi)^2 \, \sin^2 \psi = \Delta_o^2 \, \sin^2 \psi .
\end{equation}
For the computation of fluxes along the coordinate~$\rho$ it is useful to have the surface element vector~$d\vec{\Sigma}$ defined for an orthogonal coordinate system such as the oblate one expressed in terms of the variation of the position vector $\vec{r}$ along two directions $\vec{e}_1$ and $\vec{e}_2$ such that (with the indices $1$ and $2$ being $\rho$, $\psi$ or $\varphi$)
\begin{equation}\label{key}
d\vec{\Sigma} = d\vec{r}_1 \wedge d\vec{r}_2 = \frac{\partial\vec{r}}{\partial x_1} \wedge \frac{\partial\vec{r}}{\partial x_2} \, dx_1 \, dx_2 .
\end{equation}
Defined by its covariant components we get
\begin{equation}
d\Sigma_i = \sqrt{\gamma} \, \epsilon_{i12} \, dx_1 \, dx_2
\end{equation}
$\epsilon_{ijk}$ being the spatial Levi-Civita tensor \citep{arfken_mathematical_2005}. 
For instance, the radial Poynting flux across a closed surface~$\Sigma$ defined by the spheroid $\rho=\rho_0$ follows as
\begin{equation}
L = \oiint_{\Sigma} \vec{S} \cdot d\vec{\Sigma} = \oiint_{\Sigma} S^\rho \, d\Sigma_\rho
\end{equation}
with $d\Sigma_\rho = g_{\rho\rho} \, d\Sigma^\rho = \Delta_o \, \sin\psi \, d\psi \, d\varphi$ and $\vec{S}$ the Poynting vector. Exactly similar reckoning is performed for prolate coordinates as shown in th next paragraph.

\subsection{Prolate spheroidal coordinates}

The prolate spheroidal coordinate system $(\rho,\psi,\varphi)$ is given by
\begin{subequations}
	\label{eq:prolate}
	\begin{align}
	x & = \rho \, \sin \psi \, \cos\varphi \\
	y & = \rho \, \sin \psi \, \sin\varphi \\
	z & = \eta \, \cos \psi .
	\end{align}
\end{subequations}
It has the same functional form as eq.\eqref{eq:oblate} except that it inverts $\rho$ and $\eta$.
Now the polar radius of the surface delimiting the gas or the star is then $R_{\rm pol} = \sqrt{R^2+a^2}$.
The definition of the ellipticity must be changed accordingly such that 
\begin{equation}\label{eq:ellipticityProlate}
\epsilon = \frac{R_{\rm pol} - R}{(R_{\rm pol} + R)/2} .
\end{equation}
The natural basis vectors derived from eq.\eqref{eq:prolate} are expressed as
\begin{subequations}
	\label{eq:Base_naturelle_prolate}
	\begin{align}
	\eta \, \vec{e}_\rho & = \eta \, \sin \psi \, \cos \varphi \, \ex + \eta \, \sin \psi \, \sin \varphi \, \ey + \rho \, \cos \psi \, \ez \\
	\vec{e}_\psi & = \rho \, \cos \psi \, \cos \varphi \, \ex + \rho \, \cos \psi \, \sin \varphi \, \ey - \eta \, \sin \psi \, \ez \\
	\ephi & = - \rho \, \sin \psi \, \sin \varphi \, \ex + \rho \, \sin \psi \, \cos \varphi \, \ey .
	\end{align}
\end{subequations}
The position vector~$\vec{r}$ is given by
\begin{equation}
\vec{r} = \frac{\rho\,\eta^2}{\Delta_p} \, \vec{e}_\rho - \frac{a^2\,\cos\psi\,\sin\psi}{\Delta_p} \, \vec{e}_\psi
\end{equation}
with $\Delta_p = \rho^2 + a^2 \, \sin^2 \psi$.
The prolate coordinate system is also orthogonal and the metric given by
\begin{subequations}
	\begin{align}
	g_{\rho\rho} & = \frac{\rho^2 + a^2 \, \sin^2 \psi}{\eta^2} = \frac{\Delta_p}{\eta^2}\\
	g_{\psi\psi} & = \rho^2 + a^2 \, \sin^2 \psi = \Delta_p \\
	g_{\varphi\varphi} & = \rho^2 \, \sin^2 \psi .
	\end{align}
\end{subequations}
Its determinant is
\begin{equation}
\gamma = det(g) = \frac{\rho^2}{\eta^2} \,(\rho^2 + a^2 \, \sin^2 \psi)^2 \, \sin^2 \psi = \frac{\rho^2 \, \Delta_p^2}{\eta^2} \, \sin^2 \psi .
\end{equation}
The surface element on a surface $\rho=\rho_0$ is in contravariant components
\begin{equation}
	d\Sigma^\rho = \eta \, \rho \, \sin\psi \, d\psi \, d\phi .
\end{equation}

In the whole paper, we work in these coordinate systems, using the natural basis either from eq.\eqref{eq:Base_naturelle} or from eq.\eqref{eq:Base_naturelle_prolate} for the components of vector fields. We now discuss the central property of spheroidal coordinates leading to fully separable variables for the scalar Helmholtz equation.

\subsection{Separation of oblate variables}

The scalar Helmholtz equation, that reads
\begin{equation}\label{eq:helmholtz_scalaire}
\Delta W + k^2 \, W = 0 
\end{equation}
where $W$ is the unknown scalar field in three dimensions, is well known to be separable in 11~coordinate systems \citep{morse_methods_1953-1}. The spheroidal coordinates, being prolate or oblate, belong to these sets. Therefore eq.~\eqref{eq:helmholtz_scalaire} can be separated in three functions of one independent variable each, a radial part~$P(\rho)$, an angular part $\Psi(\psi)$ and an azimuthal part $\Phi(\varphi)$. The second order linear differential equations derived from this separation generate the radial and angular spheroidal wave functions \citep{olver_nist_2010, abramowitz_handbook_1965}.

Writing explicitly the solution with the ansatz $W(\rho, \psi, \varphi) = P(\rho) \, \Psi(\psi) \, \Phi(\varphi)$, the separation of variable leads to an azimuthal dependence $\Phi(\varphi) \propto e^{i\,m\,\varphi}$ where $m$ is an integer because $\Phi(\varphi)$ must be single-valued in $\varphi\in[0,2\,\pi]$ and to two second order linear differential equations for $\rho\geq0$ and $\psi\in[0,\pi]$ given respectively by
\begin{subequations}
	\begin{align}
	\label{eq:SpheroidaleRadiale}
	\frac{d}{d \rho} \left[ (\rho^2+a^2) \, \frac{d P}{d \rho} \right] + \left[ k^2 \,(\rho^2+a^2) + \frac{m^2 \, a^2}{\rho^2+a^2} \right] \, P & = + \lambda \, P \\
	\label{eq:SpheroidaleAngulaire}
	\frac{1}{\sin\psi} \, \frac{d}{d\psi} \left[\sin\psi \, \frac{d\Psi}{d\psi} \right] - \left[ k^2\, a^2\,\sin^2\psi + \frac{m^2}{\sin^2\psi} \right]\,\Psi & = -\lambda \, \Psi
	\end{align}
\end{subequations}
$\lambda$ is a separation constant, being an eigenvalue determined by the boundary conditions.
By a change to a new independent variable~$z$, letting $\rho=\pm i\,a\,z$ for eq.\eqref{eq:SpheroidaleRadiale} and $z=\cos\psi$ for eq.\eqref{eq:SpheroidaleAngulaire} both equations reduce to the same Sturm-Liouville problem
\begin{equation}
\label{eq:SpheroideEquation}
\frac{d }{d z} \left[(1-z^2) \, \frac{d f}{d z} \right] + \left[ \lambda + \gamma^2\, (1-z^2) - \frac{m^2}{1-z^2} \right]\,f = 0
\end{equation}
with $\gamma^2 = -k^2\,a^2<0$ meaning that $\gamma$ is purely imaginary. Note that the angular and radial wave functions are defined in different intervals, $|z|<1$ and $z>0$ respectively.

The most general solutions in each direction are given by
\begin{subequations}
	\begin{align}
	P(\xi) & = p_1 \, {S_\ell^m}^{(1)}(i\,\xi, \gamma) + p_2 \, {S_\ell^m}^{(2)}(i\,\xi, \gamma) \\
	\Psi(\psi) & = s_1 \, \textrm{Ps}_\ell^m(\psi, \gamma^2) + s_2 \, \textrm{Qs}_\ell^m(\psi, \gamma^2) \\
	\Phi(\varphi) & = h_1 \cos(m\,\varphi) + h_2 \sin(m\,\varphi)
	\end{align}
\end{subequations}
with $\xi=\rho/a$.
The radial ${S_\ell^m}^{(j)}$ ($j=1,2$) and angular $\textrm{Ps}_\ell^m, \textrm{Qs}_\ell^m$ spheroidal functions are defined in \cite{olver_nist_2010} and normalized according to \cite{meixner_mathieusche_1954}.
Outgoing wave solutions require to fix $P$ proportional to ${S_\ell^m}^{(3)}(i\,\xi, \gamma)$ with
\begin{equation}
{S_\ell^m}^{(3)}(i\,\xi, \gamma) = {S_\ell^m}^{(1)}(i\,\xi, \gamma) + i \, {S_\ell^m}^{(2)}(i\,\xi, \gamma)
\end{equation}
which represents a generalization of the spherical Hankel functions $h_\ell^{(1)}$ and $h_\ell^{(2)}$. The angular regularity condition imposes $s_2=0$. In complex form, a particular solution for the scalar Helmholtz equation eq.\eqref{eq:helmholtz_scalaire} with outgoing wave boundary conditions is
\begin{equation}
W = {S_\ell^m}^{(3)}(i\,\xi, \gamma) \, \textrm{Ps}_\ell^m(\psi, \gamma^2) \, e^{i\,m\,\varphi} .
\end{equation}
The same technique is applied to prolate coordinates as shown below.

\subsection{Separation of prolate variables}

Following the above lines, a same procedure for prolate coordinates leads to the angular and radial equations identical to eq.~\eqref{eq:SpheroideEquation} but with the important difference that $\gamma^2 = k^2 \, a^2>0$, meaning that $\gamma$ is real. The solution for outgoing waves now reads
\begin{equation}
W = {S_\ell^m}^{(3)}(\zeta, \gamma) \, \textrm{Ps}_\ell^m(\psi, \gamma^2) \, e^{i\,m\,\varphi}
\end{equation}
with $\zeta=\eta/a = \sqrt{1+\xi^2}$.
The arguments of ${S_\ell^m}^{(3)}$ are all real whereas for oblate coordinates they are all purely imaginary.

Rotating objects in stationary state leading to vector Helmholtz equations reducible to a set of scalar Helmholtz equations will be studied in section~\ref{sec:rotating}. However, first we need to set the background magnetic field of a static magnetized object. This is exposed in the next section.

\section{Static spheroidal star}
\label{sec:static}

For non-rotating stars, Helmholtz equation reduces to Poisson equation. By introducing a magnetic scalar potential~$\phi_M$, the solution to the magnetic field structure in vacuum can be solved like an electrostatic problem \citep{jackson_electrodynamique_2001}. We describe the procedure for any spheroidal multipole and give explicit examples of magnetic monopoles and dipoles in oblate and prolate geometries.

\subsection{Oblate magnetic star}

The magnetic field~$\vec{B}$ in vacuum outside a static star is curl free and divergenceless. It can be written as the gradient of a magnetic scalar potential~$\phi_M$ such that 
\begin{equation}\label{eq:Potentiel_Magnetic}
\vec{B} = - \vec{\nabla} \phi_M.
\end{equation}
The condition $\vec{\nabla} \cdot \vec{B} =0$ implies that the scalar potential satisfies Laplace equation 
\begin{equation}
\Delta \phi_M = 0 .
\end{equation}
This equation is fully separable in oblate spheroidal coordinates. Assuming that the inner boundary is located on the surface $\rho=\rho_0$, the general solution is expanded with $\xi=\rho/a$ into
\begin{equation}
\phi_M(\rho,\psi,\varphi) = \sum_{\ell,|m|\leq\ell} \left( a_\ell^m \, \frac{P_\ell^m(i\,\xi)}{P_\ell^m(i\,\xi_0)} + b_\ell^m \, \frac{Q_\ell^m(i\,\xi)}{Q_\ell^m(i\,\xi_0)} \right) \, Y_\ell^m(\psi,\varphi)
\end{equation}
where $P_\ell^m$ and $Q_\ell^m$ are the Legendre functions of first and second kind respectively and $Y_\ell^m$ are the spherical harmonics, see appendix~\ref{app:Legendre_functions}. The potential is imposed at $\rho=\rho_0$ with $\phi_M(\rho_0,\psi,\varphi) = V(\psi,\varphi)$ and must vanish at infinity at $\rho=+\infty$. Therefore the coefficients~$a_\ell^m$ vanish. Moreover, the coefficients $b_\ell^m$ are determined by the decomposition of the surface potential into spherical harmonics
\begin{equation}
V(\psi,\varphi) = \sum_{\ell,|m|\leq\ell} V_\ell^m \, Y_\ell^m(\psi,\varphi)
\end{equation}
and then identify $b_\ell^m = V_\ell^m$. The solution for any magnetic potential is therefore
\begin{equation}
\phi_M(\rho,\psi,\varphi) = \sum_{\ell,|m|\leq\ell} V_\ell^m \, \frac{Q_\ell^m(i\,\xi)}{Q_\ell^m(i\,\xi_0)} \, Y_\ell^m(\psi,\varphi).
\end{equation}
The magnetic field follows from Eq.~\eqref{eq:Potentiel_Magnetic}.
The physical components are
\begin{equation}
B^{\hat{i}} = - \frac{1}{\sqrt{g_{ii}}} \, \partial_i \phi_M .
\end{equation}
When the stellar shape tends to a perfect sphere, $a$ vanishes and
\begin{equation}
\lim_{a\to0}\frac{Q_\ell^m(i\,\xi)}{Q_\ell^m(i\,\xi_0)} = \left( \frac{R}{r} \right)^{\ell+1}
\end{equation}
and we retrieve the expressions for standard magnetic multipoles \citep{petri_multipolar_2015}. Note that in this limit $\rho_0 = R = R_{\rm eq}$.

\subsubsection{Monopole solution}

For completeness and to better understand the impact of the stellar ellipticity on the electromagnetic field, let us start by computing the monopole solution given by the numbers $(\ell,m) = (0,0)$. Assuming a constant potential~$V$ at its surface $\rho=R$ ($R$ should not be confused with the stellar radius that depends on colatitude~$\psi$), the magnetic potential reads
\begin{equation}
\phi_M(\rho,\psi,\varphi) = V_0^0 \, \frac{Q_0^0(i\,\xi)}{Q_0^0(i\,\xi_0)} \, Y_0^0(\psi,\varphi) = V \, \frac{\textrm{arccot }\xi}{\textrm{arccot } \xi_0}
\end{equation}
getting rid of spherical harmonic normalization factor by setting $V_0^0 = V \sqrt{4\,\pi}$.
This potential actually depends only on the coordinate~$\rho$. The magnetic field has therefore only a $\rho$ component
\begin{equation}
B_\rho = - \partial_\rho \phi_M = V \, \frac{a}{(\rho^2 + a^2) \, \textrm{arccot }\xi_0} .
\end{equation}
Introducing the constant magnetic field strength~$B$ at the surface $\rho=R$ by the definition $B=B_\rho(R)$ the magnetic surface potential reads
\begin{equation}
V = \frac{R^2+a^2}{a} \, B \, \textrm{arccot }\xi_0
\end{equation}
and the magnetic field becomes
\begin{equation}
B_\rho = B \, \frac{R^2+a^2}{\rho^2+a^2} .
\end{equation}
The physical component where $B$ is the magnetic field strength at the pole ($\rho=R,\psi=0$) is
\begin{equation}
B_{\hat \rho} = B \, \frac{R^2+a^2}{\sqrt{(\rho^2+a^2) \, (\rho^2 + a^2 \, \cos^2\psi)}} .
\end{equation}
The physical component at the equator $B_{\hat \rho}^{\rm eq}$ is stronger because
\begin{equation}
B_{\hat \rho}^{\rm eq} = B \, \sqrt{1 + \frac{a^2}{R^2}} .
\end{equation}
For $a=0$, the star becomes perfectly spherical and the field purely radial simplifies into
\begin{equation}
B_r = B \, \frac{R^2}{r^2} .
\end{equation}
In case of a small deformation with $a\ll1$ the field expands into
\begin{equation}
B_{\hat \rho} \approx B \, \frac{R^2}{\rho^2} \, \left[ 1 + \frac{a^2}{R^2} \, \left( 1 - \frac{R^2}{2\,\rho^2} \, ( 1+ \cos^2 \psi ) \right) \right] .
\end{equation}
We stress that the $\rho$ component does not correspond to the spherical radial component and as such the monopolar magnetic field has actually two component both contained in the meridional plane. In the spherical coordinate system $(r,\theta,\varphi)$, $B_\rho$ decomposes into a $B_r$ and a $B_\theta$ component because of the natural basis expressions in eq.~\eqref{eq:Base_naturelle}.

At large distances the field simplifies into
\begin{equation}
B_{\hat \rho} \approx B \, \frac{R^2}{\rho^2} \, \left[ 1 + \frac{a^2}{R^2} \right] \approx B \, \frac{R^2}{r^2} \, \left[ 1 + \frac{a^2}{R^2} \right]
\end{equation}
showing that the oblateness has still an imprint far away from the surface depicted by the correcting factor $(1+a^2/R^2)$. This factor corresponds to an increase in the magnetic field strength compared to a spherical dipole with surface field~$B$ as measured by a distant observer.

\subsubsection{Dipole solution}

The procedure to follow for the dipole or for any multipole is very similar to the monopole case. Because of the linearity of the problem, we solve separately for an aligned and an orthogonal dipole. Even in the static case, both solutions are different because the ellipsoid defines new preferred axes breaking the spherical symmetry.

For an oblique rotator, the magnetic potential with parallel~$V_1^0$ and perpendicular~$V_1^1$ contribution is
\begin{equation}
\phi_M(\rho,\psi,\varphi) = V_1^0 \, \frac{Q_1^0(i\,\xi)}{Q_1^0(i\,\xi_0)} \, Y_1^0(\psi,\varphi) + V_1^1 \, \frac{Q_1^1(i\,\xi)}{Q_1^1(i\,\xi_0)} \, Y_1^1(\psi,\varphi)
\end{equation}
or again getting rid of spherical harmonic normalization factors with parallel~$V_\parallel$ and perpendicular~$V_\perp$ contributions and a possible negative sign for the magnetic field components we write
\begin{multline}
- \phi_M(\rho,\psi,\varphi) =  V_\parallel \, \frac{\xi \, \textrm{arccot }\xi - 1}{\xi_0 \, \textrm{arccot }\xi_0 - 1} \, \cos \psi +  \\ V_\perp \, \frac{(\xi^2+1) \, \textrm{arccot }\xi - \xi}{(\xi_0^2+1) \, \textrm{arccot }\xi_0 - \xi_0} \, \sqrt{\frac{\xi_0^2+1}{\xi^2+1}} \sin\psi \, e^{i\, \varphi }.
\end{multline}
The magnetic field is decomposed into an aligned rotator as
\begin{subequations}
\begin{align}
B_\rho & = \frac{V_\parallel}{a} \, \frac{(\xi^2+1) \, \arccot \xi - \xi}{(\xi^2+1) \, (\xi_0 \, \arccot \xi_0 - 1)} \, \cos \psi \\
B_\psi & = - V_\parallel \, \frac{\xi \, \arccot \xi- 1 }{\xi_0 \, \arccot \xi_0 - 1} \, \sin \psi \\
B_\varphi & = 0
\end{align}
\end{subequations}
or by introducing a typical surface magnetic field strength~$B_\parallel = B_\rho(R,\psi=0)$
\begin{subequations}
	\begin{align}
	B_\rho & = 2 \, B_\parallel\, \frac{\xi_0^2+1}{\xi^2+1} \, \frac{(\xi^2+1)\,\arccot \xi - \xi}{(\xi_0^2+1) \, \arccot \xi_0 - \xi_0} \, \cos \psi \\
	B_\psi & = 2 \, a \, B_\parallel\, \frac{(\xi_0^2+1) \, (1 - \xi \, \arccot \xi)}{(\xi_0^2+1) \, \arccot \xi_0 - \xi_0} \, \sin \psi \\
	B_\varphi & = 0
	\end{align}
\end{subequations}
and an orthogonal rotator as
\begin{subequations}
	\begin{align}
	B_\rho & = \frac{V_\perp}{a} \, \frac{(\xi^2+1) \, \xi \, \arccot \xi - \xi^2 - 2}{(\xi_0^2+1) \, \arccot \xi_0 - \xi_0} \, \frac{\sqrt{\xi_0^2+1}}{(\xi^2+1)^{3/2}} \, \sin \psi \, e^{i\, \varphi }\\
	B_\psi & = V_\perp \, \frac{(\xi^2+1) \, \textrm{arccot }\xi - \xi}{(\xi_0^2+1) \, \textrm{arccot }\xi_0 - \xi_0} \, \sqrt{\frac{\xi_0^2+1}{\xi^2+1}} \, \cos \psi \, e^{i\, \varphi } \\
	B_\varphi & = i\, V_\perp \, \frac{(\xi^2+1) \, \textrm{arccot }\xi - \xi}{(\xi_0^2+1) \, \textrm{arccot }\xi_0 - \xi_0} \, \sqrt{\frac{\xi_0^2+1}{\xi^2+1}} \, \sin \psi \, e^{i\, \varphi } .
	\end{align}
\end{subequations}
or by introducing a typical surface magnetic field strength~$B_\perp = B_\rho(R,\psi=\pi/2,\varphi=0)$
	\begin{subequations}
	\begin{align}
	B_\rho & = 2 \, B_\perp \, \frac{\xi^2 + 2 - (\xi^2+1) \, \xi \, \arccot \xi}{\xi_0^2 + 2 - (\xi_0^2+1) \, \xi_0 \, \arccot \xi_0} \, \left( \frac{\xi_0^2+1}{\xi^2+1}\right)^{3/2} \, \sin \psi \, e^{i\, \varphi }\\
	B_\psi & = 2 \, a \, B_\perp \, \frac{(\xi^2+1) \, \arccot \xi - \xi}{(\xi_0^2+1) \, \xi_0 \, \arccot \xi_0 - \xi_0^2 - 2} \, \frac{(\xi_0^2+1)^{3/2}}{\sqrt{\xi^2+1}} \, \cos \psi \, e^{i\, \varphi } \\
	B_\varphi & = 2 \, i \, a \, B_\perp \, \frac{(\xi^2+1) \, \arccot \xi - \xi}{(\xi_0^2+1) \, \xi_0 \, \arccot \xi_0 - \xi_0^2 - 2} \, \frac{(\xi_0^2+1)^{3/2}}{\sqrt{\xi^2+1}} \, \sin \psi \, e^{i\, \varphi } .
	\end{align}
\end{subequations}
In order to better connect these expressions to the spherical dipole, we introduced the magnetic field strength at the north pole by respectively $2\,B_\parallel$ and $2\,B_\perp$ for the aligned and orthogonal rotator. Because the metric coefficient~$g_{\rho\rho}=1$ at the pole for an oblate coordinate system, the natural component is equal to the physical component, therefore $B_{\hat \rho}  = B_\rho = 2\,  B_{\parallel}$ for aligned and $B_{\hat \rho}  = B_\rho = 2\,  B_{\perp}$ for perpendicular rotators.

At large distances, aligned and perpendicular components of an oblate dipole tend respectively to
\begin{subequations}
	\begin{align}
	B_\rho & = \frac{4}{3} \, B_\parallel \, \frac{a^3}{\rho^3} \, \frac{\xi_0^2+1}{(\xi_0^2 + 1) \, \arccot \xi_0 - \xi_0} \, \cos \psi \\
	B_\psi & = \frac{2}{3} \, B_\parallel \, \frac{a^3}{\rho^2} \, \frac{\xi_0^2+1}{(\xi_0^2 + 1) \, \arccot \xi_0 - \xi_0} \,\sin \psi \\
	B_\varphi & = 0
	\end{align}
\end{subequations}
for the aligned part and
\begin{subequations}
	\begin{align}
	B_\rho & = \frac{8}{3} \, B_\perp \, \frac{a^3}{\rho^3} \,\frac{(\xi_0^2+1)^{3/2}}{\xi_0^2 + 2 - (\xi_0^2+1) \, \xi_0 \, \arccot \xi_0} \, \sin \psi \, e^{i\, \varphi }\\
	B_\psi & = \frac{4}{3} \, B_\perp \, \frac{a^3}{\rho^2} \,\frac{(\xi_0^2+1)^{3/2}}{(\xi_0^2+1) \, \xi_0 \, \arccot \xi_0 - \xi_0^2 - 2} \,  \cos \psi \, e^{i\, \varphi } \\
	B_\varphi & = \frac{4}{3} \, i\, B_\perp \, \frac{a^3}{\rho^2} \,\frac{(\xi_0^2+1)^{3/2}}{(\xi_0^2+1) \, \xi_0 \, \arccot \xi_0 - \xi_0^2 - 2} \, \sin \psi \, e^{i\, \varphi }
	\end{align}
\end{subequations}
for the orthogonal part.


\subsection{Prolate magnetic star}

Following the same procedure as for the oblate magnetic star we find the magnetic potential $\phi_M$ for prolate star as
\begin{equation}
\phi_M(\rho,\psi,\varphi) = \sum_{\ell,|m|\leq\ell} V_\ell^m \, \frac{Q_\ell^m(\zeta)}{Q_\ell^m(\zeta_0)} \, Y_\ell^m(\psi,\varphi)
\end{equation}
with $\zeta = \sqrt{1+(\rho/a)^2}$ and  $\zeta_0 = \sqrt{1+(R/a)^2}$.

For small prolateness tending to a spherical shape, $a$ tends to zero and the potential simplifies to a multipole like
\begin{equation}
\lim_{a\to0}\frac{Q_\ell^m(\zeta)}{Q_\ell^m(\zeta_0)} = \left( \frac{R}{r} \right)^{\ell+1}
\end{equation}
the same expression as in the previous subsection because oblate and prolate tend both to a spherical shape when $a\to0$.

\subsubsection{Monopole solution}

For the monopole, the potential reads explicitly
\begin{equation}
\phi_M(\rho,\psi,\varphi) = V_0^0 \, \frac{Q_0^0(\zeta)}{Q_0^0(\zeta_0)} \, Y_0^0(\psi,\varphi) = V \, \frac{\textrm{arccoth }\zeta}{\textrm{arccoth } \zeta_0}.
\end{equation}
It actually depends only on the $\rho$ coordinate. The magnetic field has therefore only a $\rho$ component
\begin{equation}
B_\rho = - \partial_\rho \phi_M = V \, \frac{a}{\rho \, \sqrt{\rho^2 + a^2} \, \textrm{arccoth }\zeta_0} .
\end{equation}
At the stellar surface $\rho=R$ and $B_\rho=B$ thus
\begin{equation}
V = \frac{R \, \sqrt{R^2 + a^2}}{a} \, B \, \textrm{arccoth }\zeta_0
\end{equation}
such that
\begin{equation}
B_\rho = B \, \frac{R}{\rho} \sqrt{ \frac{R^2+a^2}{\rho^2+a^2} }.
\end{equation}
The physical component where $B$ is the magnetic field strength at the equator ($\rho=R,\psi=\pi/2$) is
\begin{equation}
B_{\hat \rho} = B\, \frac{R}{\rho} \, \sqrt{ \frac{R^2+a^2}{\rho^2 + a^2 \sin^2\psi} } .
\end{equation}
The physical component at the pole $B_{\hat \rho}^{\rm pole}$ is stronger because
\begin{equation}
	B_{\hat \rho}^{\rm pole} = B \, \sqrt{1 + \frac{a^2}{R^2}} .
\end{equation}
For small prolateness $a\ll R$, the potential and the field becomes
\begin{subequations}
\begin{align}
\phi_M & \approx V \, \frac{R}{\rho} \, \left[ 1 + \frac{1}{6} \, \frac{a^2}{R^2} \, \left( 1 - \frac{R^2}{\rho^2} \right) \right] \\
B_{\hat \rho} & \approx B \, \frac{R^2}{\rho^2} \, \left[ 1 + \frac{1}{2} \, \frac{a^2}{R^2} \, \left( 1 - \frac{R^2}{\rho^2} \sin^2\psi \right) \right]
\end{align}
\end{subequations}
reducing to the spherical monopole in the limit of a perfect sphere.

At large distances, the  physical component reduces to
\begin{equation}
B_{\hat \rho} = B\, \frac{R^2}{\rho^2} \, \sqrt{ 1 + \frac{a^2}{R^2} } .
\end{equation}
Here also we observe a correcting factor but of value $\sqrt{ 1 + a^2/R^2 }$ compared to a spherical monopole.

\subsubsection{Dipole solution}

For the prolate dipole, the potential with parallel~$V_1^0$ and perpendicular~$V_1^1$ contribution is
\begin{equation}
\phi_M(\rho,\psi,\varphi) = V_1^0 \, \frac{Q_1^0(\zeta)}{Q_1^0(\zeta_0)} \, Y_1^0(\psi,\varphi) + V_1^1 \, \frac{Q_1^1(\zeta)}{Q_1^1(\zeta_0)} \, Y_1^1(\psi,\varphi).
\end{equation}
or explicitly with parallel~$V_\parallel$ and perpendicular~$V_\perp$ contributions (getting rid of spherical harmonic normalization factors)
\begin{multline}
	- \phi_M(\rho,\psi,\varphi) =  V_\parallel \, \frac{\zeta \, \textrm{arccoth }\zeta - 1}{\zeta_0 \, \textrm{arccoth }\zeta_0 - 1} \, \cos \psi + \\ V_\perp \, \frac{(\zeta^2-1) \, \textrm{arccoth }\zeta - \zeta}{(\zeta_0^2-1) \, \textrm{arccoth }\zeta_0 - \zeta_0} \, \sqrt{\frac{\zeta_0^2-1}{\zeta^2-1}} \, \sin\psi \, e^{i\, \varphi } .
\end{multline}
As for oblate stars in vacuum, because of linearity, we solve separately for an aligned and an orthogonal rotator. For the aligned rotator, the magnetic field is
\begin{subequations}
	\begin{align}
		B_\rho & = - \frac{V_\parallel}{\rho} \left[ 1- \frac{\rho^2}{a^2} \, \frac{\textrm{arccoth }\zeta}{\zeta}  \right] \, \frac{1}{\zeta_0 \, \textrm{arccoth }\zeta_0 - 1} \, \cos \psi \\
		B_\psi & = V_\parallel \, \frac{1 - \zeta \, \textrm{arccoth }\zeta}{\zeta_0 \, \textrm{arccoth }\zeta_0 - 1} \, \sin \psi \\
		B_\varphi & = 0 .
	\end{align}
\end{subequations}
By introducing a typical magnetic field strength at the surface, we get
\begin{subequations}
	\begin{align}
	B_\rho & = 2 \, B_\parallel \, \frac{R}{\rho} \, \left[ 1- \xi^2 \, \frac{\textrm{arccoth }\zeta}{\zeta}  \right] \, \left[ 1- \xi_0^2 \, \frac{\textrm{arccoth }\zeta_0}{\zeta_0}  \right]^{-1} \, \cos \psi \\
	B_\psi & = 2 \, B_\parallel \, R \, \left[ \zeta \, \textrm{arccoth }\zeta - 1 \right] \, \left[ 1- \xi_0^2 \, \frac{\textrm{arccoth }\zeta_0}{\zeta_0}  \right]^{-1} \, \sin \psi \\
	B_\varphi & = 0 .
	\end{align}
\end{subequations}
For the orthogonal rotator, we found
\begin{subequations}
	\begin{align}
		B_\rho & = \frac{V_\perp}{a} \, \frac{\sqrt{\zeta_0^2 - 1} \, ((\zeta^2-1) \, \zeta \, \textrm{arccoth } \zeta - \zeta^2 + 2 )}{((\zeta_0^2-1) \, \textrm{arccoth }\zeta_0 - \zeta_0) \, (\zeta^2-1) \, \zeta}  \, \sin \psi \, e^{i\, \varphi }\\
		B_\psi & = V_\perp \,  \frac{(\zeta^2-1) \, \textrm{arccoth }\zeta - \zeta}{(\zeta_0^2-1) \, \textrm{arccoth }\zeta_0 - \zeta_0} \, \sqrt{\frac{\zeta_0^2-1}{\zeta^2-1}} \,  \cos \psi \, e^{i\, \varphi } \\
		B_\varphi & = i\,V_\perp \, \frac{(\zeta^2-1) \, \textrm{arccoth }\zeta - \zeta}{(\zeta_0^2-1) \, \textrm{arccoth }\zeta_0 - \zeta_0} \, \sqrt{\frac{\zeta_0^2-1}{\zeta^2-1}} \sin\psi \, e^{i\, \varphi }
	\end{align}
\end{subequations}
or with the typical surface magnetic field strength~$B_\perp$
\begin{subequations}
	\begin{align}
	B_\rho & = 2 \, B_\perp \, \frac{\zeta_0}{\zeta} \,\frac{\zeta_0^2 - 1}{\zeta^2 - 1} \,\frac{(\zeta^2-1) \,\zeta \,  \textrm{arccoth } \zeta - \zeta^2 +2}{(\zeta_0^2-1) \, \zeta_0 \, \textrm{arccoth }\zeta_0 - \zeta_0^2 + 2}   \, \sin \psi \, e^{i\, \varphi }\\
	B_\psi & = 2 \, a \, B_\perp \, \frac{\zeta_0\,(\zeta_0^2-1)}{\sqrt{\zeta^2-1}} \frac{(\zeta^2-1) \, \textrm{arccoth }\zeta - \zeta}{(\zeta_0^2-1) \, \zeta_0 \, \textrm{arccoth }\zeta_0 - \zeta_0^2 + 2} \,  \cos \psi \, e^{i\, \varphi } \\
	B_\varphi & = 2 \, i\, a \, B_\perp \,  \frac{\zeta_0\,(\zeta_0^2-1)}{\sqrt{\zeta^2-1}} \frac{(\zeta^2-1) \, \textrm{arccoth }\zeta - \zeta}{(\zeta_0^2-1) \, \zeta_0 \, \textrm{arccoth }\zeta_0 - \zeta_0^2 + 2} \, \sin\psi \, e^{i\, \varphi.}
	\end{align}
\end{subequations}
At large distances, for the aligned component we get
\begin{subequations}
	\begin{align}
	B_\rho & = \frac{4}{3} \, B_\parallel \, \frac{a^3}{\rho^3} \, \frac{\zeta_0 \, \sqrt{\zeta_0^2-1}}{\zeta_0 + (1 -\zeta_0^2 ) \, \textrm{arccoth }\zeta_0} \, \cos \psi \\
	B_\psi & = \frac{2}{3} \, B_\parallel \, \frac{a^3}{\rho^2} \, \frac{\zeta_0 \, \sqrt{\zeta_0^2-1}}{\zeta_0 + (1 -\zeta_0^2 ) \, \textrm{arccoth }\zeta_0} \, \sin \psi \\
	B_\varphi & = 0 .
	\end{align}
\end{subequations}
and for the orthogonal component
\begin{subequations}
	\begin{align}
	B_\rho & = \frac{8}{3} \, B_\perp \, \frac{a^3}{\rho^3} \, \frac{\zeta_0 \, (\zeta_0^2-1) \, \sin \psi \, e^{i\, \varphi }}{(\zeta_0^2-1) \, \zeta_0 \, \textrm{arccoth }\zeta_0 - \zeta_0^2 + 2} \\
	B_\psi & = \frac{4}{3} \, B_\perp \, \frac{a^3}{\rho^2} \, \frac{\zeta_0 \, (\zeta_0^2-1) \,\cos \psi \, e^{i\, \varphi }}{\zeta_0^2 - 2 -(\zeta_0^2-1) \, \zeta_0 \, \textrm{arccoth }\zeta_0} \\
	B_\varphi & = \frac{4}{3} \, i\,B_\perp \, \frac{a^3}{\rho^2} \, \frac{\sin\psi \, e^{i\, \varphi }}{\zeta_0^2 - 2 - (\zeta_0^2-1) \, \zeta_0 \, \textrm{arccoth }\zeta_0} .
	\end{align}
\end{subequations}
This achieves the implementation of the initial background magnetic field set up to start the numerical simulations in section~\ref{sec:simulations}. A last important topic concerns the field normalization convention used to compare results obtained with different neutron star geometries.



\subsection{Field normalization}

Our main goal is to compare spheroidal stars to perfect spherical stars by computing some relevant quantities like for instance the Poynting flux. An important related question concerns the normalization of the spheroidal field compared to the corresponding spherical case. The impact of the stellar shape will drastically influence these quantities. Therefore we need a reference configuration with an appropriately chosen magnetic field strength at the surface. However there exists obviously several approaches to fix the electromagnetic field at the surface like keeping a fixed value of the magnetic field strength at the pole or at the equator when deforming the stellar surface. Nevertheless this seems not satisfactorily because the spin down luminosity for instance varies not only because of the spheroidal shape but also because of the artificial field strength variation related to the evolving boundary. This problem is reminiscent to the normalization of the magnetic dipole or multipole in a curved space time of general relativity. There the normalization at the surface is chosen in order to keep the asymptotic structure at large distances identical whatever the compactness and frame dragging effects. We decided to use the same techniques to normalize the surface spheroidal field, imposing a dipole magnetic field at large distances tending always to the same perfect spherical dipole keeping a constant asymptotic expression. Would we normalize it differently, the estimate of the spin down would change significantly. With our procedure, we expect to minimize all variations imputed to the field strength value at the surface retaining only the true effect of spheroidal shapes. This normalization must be carefully exposed in order to compare with previous results like for instance \cite{finn_spin-down_1990} who assumed a star keeping a constant magnetic flux and not a constant asymptotic magnetic dipole moment.

\section{Rotating spheroidal stars}
\label{sec:rotating}

The quasi-stationary solution is acceptable for slowly rotating stars or when looking in regions well inside the light-cylinder. However when seeking for the field behaviour at large distances, it switches to a wave nature around the light-cylinder due to the finite speed of propagation of electromagnetic fields~$\mathbf{F}$. In this section, we solve for these fields in the whole space outside the star. Both the electric and the magnetic part satisfy the vector wave equation in vacuum given by
\begin{equation}
\frac{1}{c^2} \, \frac{\partial^2 \mathbf{F}}{\partial t^2} - \Delta \mathbf{F} = \mathbf{0} .
\end{equation}
For periodic motion, the time dependence becomes harmonic and the field~$\mathbf{F}$ varies in time according to $\mathbf{F} \propto e^{-i\,\omega\,t}$. Introducing the wave number $k=\omega/c$, the vector wave equation reduces to the vector Helmholtz equation
\begin{equation}\label{eq:helmholtz_vecteur}
 \Delta \mathbf{F} + k^2 \, \mathbf{F} = \mathbf{0} .
\end{equation}
Next we show how to find exact analytical solutions in spheroidal coordinates at least formally.

\subsection{Time harmonic solutions}

Before treating the vector Helmholtz equation, it is useful to remind the results about the scalar Helmholtz equation~\eqref{eq:helmholtz_scalaire}. It can be shown that if $W$ is a solution of \eqref{eq:helmholtz_scalaire} then $\vec{\nabla}W$, $\vec{\Phi} = \vec{\nabla} \wedge (W \, \vec{r})$ and $\vec{\Psi} = \vec{\nabla} \wedge \vec{\Phi}$ are all solutions of the vector Helmholtz equation~\eqref{eq:helmholtz_vecteur} \citep{leitner_oblate_1950, gumerov_fast_2004}. Moreover, $\vec{\Phi}$ and $\vec{\Psi}$ are solenoidal, meaning that $\vec{\nabla} \cdot \vec{\Phi} = \vec{\nabla} \cdot \vec{\Psi} = 0$ and they satisfy $\vec{\nabla} \wedge \vec{\Psi} = k^2 \, \vec{\Phi}$.

The time harmonic vacuum Maxwell equations can then be solved by expanding the electric and magnetic field into \citep{asano_light_1975}
\begin{subequations}
\begin{align}
\vec{E} & = \sum_{\ell, m} a^{\rm E}_{\ell m} \, \vec{\Psi}_{\ell m} + b^{\rm E}_{\ell m} \, \vec{\Phi}_{\ell m} \\
c\,\vec{B} & = \sum_{\ell, m} a^{\rm B}_{\ell m} \, \vec{\Psi}_{\ell m} + b^{\rm B}_{\ell m} \, \vec{\Phi}_{\ell m}
\end{align}
\end{subequations}
automatically satisfying $\vec{\nabla} \cdot \vec{E} = \vec{\nabla} \cdot \vec{B} = 0$. The numbers $\ell$ and $m$ label the multipole expansion modes similarly to the vector spherical harmonics \citep{petri_general-relativistic_2013}. The coefficients $a^{\rm E/B}_{\ell m}$ and $b^{\rm E/B}_{\ell m}$ are constants of integration depending on the boundary conditions. In order to satisfy the time-harmonic Maxwell equations with temporal behaviour $e^{-i\,\omega\,t}$, these coefficients must be related by
\begin{subequations}
	\begin{align}
	c\,b^{\rm E}_{\ell m} & = +i\,\omega\, a^{\rm B}_{\ell m} \\
	c\, b^{\rm B}_{\ell m} & = -i\,\omega\, a^{\rm E}_{\ell m} .
	\end{align}
\end{subequations}
Full solutions to Maxwell equations in vacuum are therefore summarized by the expansion
\begin{subequations}
	\begin{align}
	\vec{E} & = \sum_{\ell, m} a^{\rm E}_{\ell m} \, \vec{\Psi}_{\ell m} + i\,k\, a^{\rm B}_{\ell m} \, \vec{\Phi}_{\ell m} \\
	c\,\vec{B} & = \sum_{\ell, m} a^{\rm B}_{\ell m} \, \vec{\Psi}_{\ell m} -i\,k\, a^{\rm E}_{\ell m} \, \vec{\Phi}_{\ell m} .
	\end{align}
\end{subequations}
The only independent coefficients are therefore $a^{\rm E/B}_{\ell m}$. The central problem is to find explicit expressions for the vector $\vec{\Psi}_{\ell m}$ and $\vec{\Phi}_{\ell m}$ in spheroidal coordinates and to adjust the coefficients $a^{\rm E/B}_{\ell m}$ to fit the stellar boundary conditions.


\subsection{Electric field}

As often done for neutron star interiors, we assume a perfect conductor inside the star leading to an electric field in the observer frame given by $\vec{E} + \vec{v} \wedge \vec{B} = \vec{0}$ where 
\begin{equation}\label{eq:MHD_ideale}
\vec{v} = \vec{\Omega} \wedge \vec{r}
\end{equation}
is the rotation speed of the star. Outside the star, the electric field is divergenceless as the magnetic field. Adapted to the spheroidal coordinates, the radial component~$E^\rho$ is unconstrained, $E^\varphi=0$ and 
\begin{equation}\label{eq:boundary}
E^\psi = - \sqrt{\frac{g_{\rho\rho} \, g_{\varphi\varphi}}{g_{\psi\psi}}} \, \Omega \, B^\rho.
\end{equation}
These boundary conditions on the electric field completely and uniquely define the whole electromagnetic field in vacuum.

\subsection{Approximate solution for oblique rotators}

There are no exact analytical closed forms for the solutions presented above. However, the parameter $\gamma^2 = \pm k^2\,a^2$ remains always less than one in modulus because the star must remain within its light cylinder thus the constrain $k\,a = a/\rlight\ll1$. Therefore we can expand the solution into a series in $\gamma$ that converges quickly to the exact expression. In this section we follow this path to get more insight into the impact of oblateness/prolateness on the Poynting flux. Our starting point are the expansion formulae for the angular and radial spheroidal wave functions as given by \cite{abramowitz_handbook_1965}. We summarize the important results useful for our reckoning. The oblate geometry is tightly related to the prolate geometry and the results are often only given for prolate functions. Switching to oblate coordinates only requires a change of variables such that $\rho \rightarrow \pm i \, \rho$ and $\gamma \rightarrow \mp i \, \gamma$. For brevity we only give results for prolate shapes.

The prolate angular spheroidal wave functions of the first kind are expanded into Legendre functions of the first kind via
\begin{equation}
\textrm{Ps}_\ell^m(\gamma,\psi) = {\sum_{r=0,1}^{+\infty}}' d_r^{m\ell} \, P_{m+r}^m(\psi)
\end{equation}
where the prime indicates summation from 0/1 over even/odd indices when $(\ell-m)$ is even/odd.
The expansion coefficients $d_r^{m\ell}$ are determined by solving a three-term recurrence relation with coefficients given in \cite{abramowitz_handbook_1965}. 

The prolate radial wave functions are then expanded into
\begin{equation}
S_\ell^m(\gamma, \zeta) = \frac{\left( 1 - \zeta^{-2}\right)^{m/2}}{{\sum_{r=0,1}^{+\infty}}' d_r^{m\ell} \, \frac{(2\,m+r)!}{r!}} \, {\sum_{r=0,1}^{+\infty}}' d_r^{m\ell} \, \frac{(2\,m+r)!}{r!} \, i^{r+m-\ell} \, z_{m+r}(\gamma\,\zeta)
\end{equation}
where $z_{\ell}$ is any of the spherical Bessel function $j_\ell,y_\ell$ or spherical Hankel function $h_\ell^{(1)}$ or $h_\ell^{(2)}$. For our problem of outgoing wave solutions, we require a radial expansion into spherical Hankel function of type $h_\ell^{(1)}$ as for the Deutsch solution.

For analytically tractable purpose, we expand all parameters to second order in $\gamma$. Actually, the coefficients $d_r^{m\ell}$ depend only on even powers of $\gamma$ therefore the expansion of any quantity will also follow an expansion in even powers of $\gamma$. Consequently, the first correcting term for magnetic field structure, Poynting flux, electromagnetic torque and so on will depend on $\gamma^2$. The key expansion coefficients are remind in appendix~\ref{app:Angular_functions_expansion}.

\subsection{Dipole radiation}

Unfortunately the eigenvector expansion in spheroidal coordinates does not allow an identification term by term of each mode $(\ell,m)$ as would be possible in spherical coordinates. To get more analytical insight into the Poynting flux perturbed by the shape, we need to resort to a series expansion of the eigenvectors. We identify various contributions to the Poynting flux, the magnetic dipole being the dominant loss channel. For a spherical star, the rotating magnetic dipole induces a quadrupole electric field that also radiates. Nevertheless, this quadrupole brings in corrections to a point dipole that are of much higher order in spin parameter $w=\Omega\,R/c$, at least $w^4$ compared to $1$ and $w^2$ for the dipole. This is due to the fact that the quadrupole is already the result of rotating a magnetic field, thus an $w^2$ strength for a $w^2$ spin down rate. 

We expect this assertion to hold for a spheroidal star, meaning that useful and exact corrections can be found solely by computing the magnetic radiation part to second order in~$w$ without contributions from the electric quadrupole. Interesting results are then derived by solving for the coefficient $a^{\rm B}_{1,1}$ only. The Poynting flux in such an approximation then behaves like
\begin{equation}
L \approx \frac{c}{2\,\mu_0} \, |a^{\rm B}_{1,1}|^2 .
\end{equation}
The dominant term in the magnetic dipole radiation is given by the model $(\ell,m)=(1,1)$. The prolate radial wave function is therefore to second order in~$\gamma$ given by
\begin{equation}
S_1^1(\gamma, \zeta) \approx \left[ 1 - \left( \frac{2 \, \gamma^2}{25} + \frac{a^2}{2\,\rho^2} \right) \right] \, h_1^{(1)}(k\,\rho) - \frac{2}{25} \,  \gamma^2 \, h_3^{(1)}(k\,\rho) + O(\gamma^4)
\end{equation}
For outgoing wave boundary conditions we have set $z_\ell=h_\ell^{(1)}$. Following the procedure explained in detail in \cite{petri_multipolar_2015}, the spin down dependence on $a/R$ and $R/\rlight$ is approximately proportional to $|S_1^1(\gamma, \zeta)|^{-2}$. An expansion to lowest order in these two parameters gives a correction to the luminosity as
\begin{equation}\label{eq:Lprolate}
 L_{\rm prolate} \approx L_\perp \, \left[1 - \left( \frac{R}{\rlight} \right)^2 + \frac{27}{5} \, \left( \frac{a}{R} \right)^2 - \frac{36}{5} \,  \left( \frac{a}{\rlight} \right)^2  \right] 
\end{equation}
where the spin down~$L_\perp$ of the vacuum orthogonal point magnetic dipole is
\begin{equation}\label{eq:Lperp}
L_\perp = \frac{8\,\pi}{3\,\mu_0\,c^3}\,\Omega^4\,B^2\,R^6 .
\end{equation}
The above approximation gives only some important hints about the spin down change due to the spheroidal shape. It does not take into account the normalization of the surface magnetic field strength.
Similar analytical investigation can be performed for an oblate star, but contrary to vector spherical harmonics, vector spheroidal harmonics as defined in this work do not permit to impose the stellar surface boundary conditions on the electric field in a closed analytical form because even though the scalar spheroidal harmonics naturally embrace the spheroidal shape, their vector counterpart do not decompose easily into tangential and normal component on a spheroidal object. Therefore, imposing continuity of the normal component of the magnetic field and continuity of the tangential component of the electric field is a non trivial task. Nevertheless, it clearly shows that leading corrections scale as $\gamma^2 = k^2\,a^2 = a^2/\rlight^2$ and $a^2/R^2$, thus are of second order in $a$. We will use this results to fit the numerical simulations performed in the next section.

\section{Time-dependent simulations}
\label{sec:simulations}

Analytically solving the Helmholtz equation with separate coordinates helps to get insight into the effect of oblateness or prolateness on the spin-down luminosity and magnetic field structure. However, the boundary conditions on the stellar surface cannot be imposed with a finite number of terms in angular spheroidal wave functions contrary to spherical harmonics for perfect spheres. It is therefore enlightening to compute numerical solutions by performing time-dependent simulations of Maxwell equations in vacuum by properly taking into account the boundary conditions on the surface with high accuracy to catch the effect of the surface electric field. This last section presents the results of such computations, first showing the structure of field lines, then investigating the spin-down luminosity and eventually tracing the shape of the polar caps.

\subsection{Numerical setup}

Maxwell equations are solved with our pseudo-spectral code developed in \cite{petri_pulsar_2012}. However in order to better resolve the inner computational domain, we map the usual Chebyshev grid to a truncated rational Chebyshev grid, increasing the resolution in the inner part with respect to the outer part \citep{boyd_chebyshev_2001}. This allows us to use a coarser grid of only $N_r \times N_\theta \times N_\varphi = 129\times32\times64$. The neutron star is a perfect conductor imposing an electric field on its surface given by Eq.~\eqref{eq:boundary}. The neutron star radius is set to $R/\rlight=0.3$ which coincides with the inner boundary of the computational domain $R_1=R$. The outer boundary is equal to $R_2/\rlight=7$. The oblateness or prolateness is controlled by the parameter~$a$ defining $\eta$ in spheroidal coordinates. This parameter~$a/R$ spans the range $[0,1]$ although it is not bounded by $R$ but by the fact that the equatorial radius of the star cannot exceed the light-cylinder radius. The obliquity~$\rchi$ is taken in the set $\rchi \in \{0\degr, 30\degr, 60\degr, 90\degr\}$.

\subsection{Magnetic field lines}

\begin{figure*}
	\centering
	\includegraphics[width=0.95\linewidth]{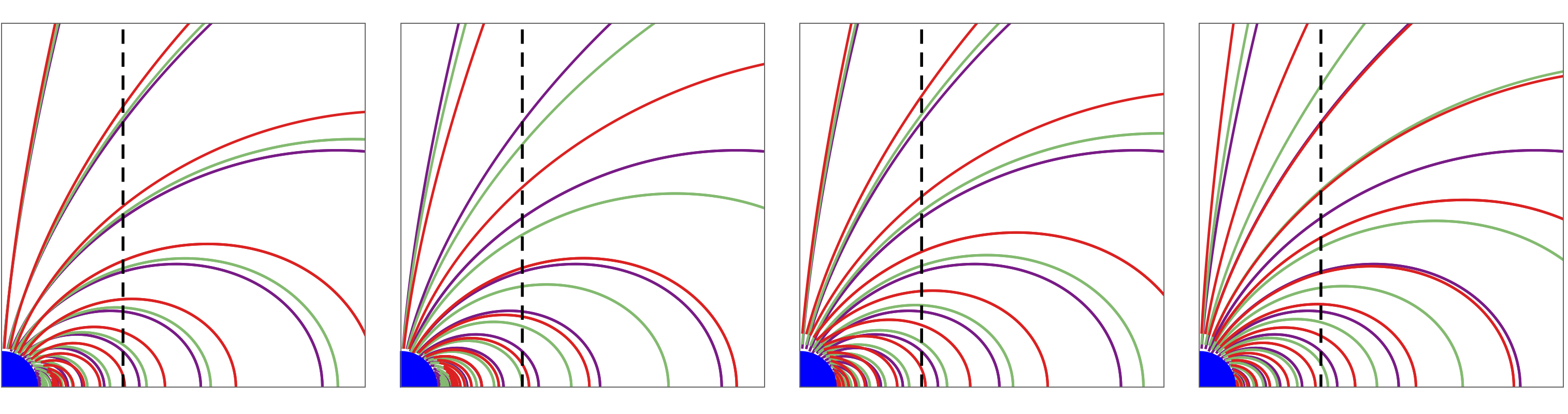}
	\caption{Magnetic field lines in the meridional plane $xOz$ for an aligned spheroidal rotator with oblateness (first two columns) or prolateness (last two columns) parameter $a/R=\{0,0.5,1\}$ respectively in blue, green and red. The blue disk on the bottom left represents the spherical star.}
	\label{fig:lignes_champ_B_xz_a0}
\end{figure*}

\begin{figure*}
	\centering
	\includegraphics[width=0.95\linewidth]{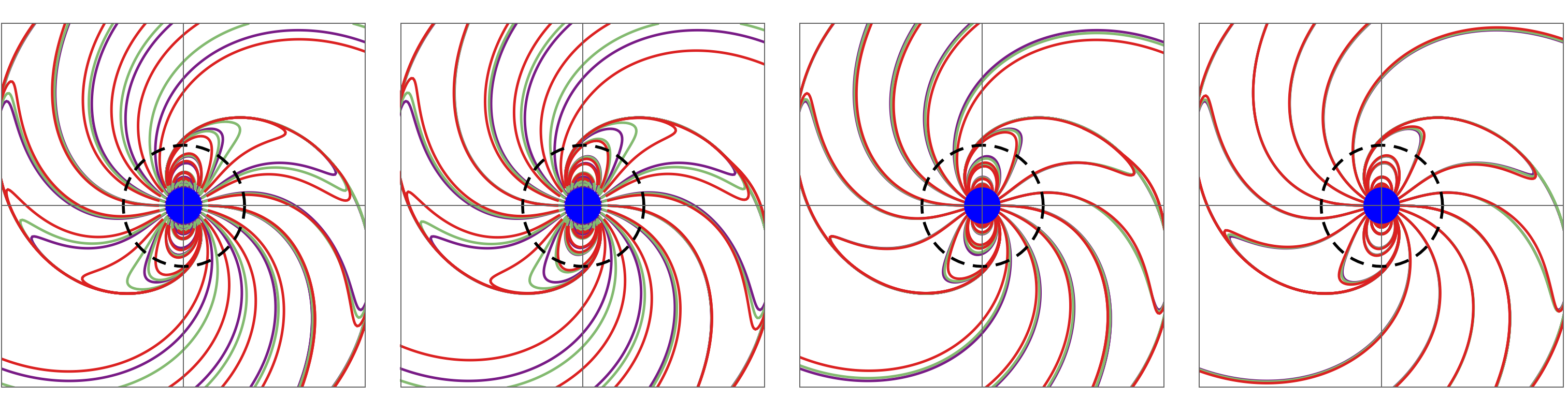}
	\caption{Magnetic field lines in the equatorial plane $xOy$ for a perpendicular spheroidal rotator with oblateness (first two columns) or prolateness (last two columns) parameter $a/R=\{0,0.5,1\}$ respectively in blue, green and red.}
	\label{fig:lignes_champ_B_xy_a90}
\end{figure*}

Let us first compare the impact of the spheroidal shape onto the magnetic field line structure for an aligned and a perpendicular rotator for ease to plot accurately in 2D. They are shown respectively in Fig.~\ref{fig:lignes_champ_B_xz_a0} and Fig.~\ref{fig:lignes_champ_B_xy_a90} for oblate and prolate stars with different boundary conditions, either single multipolar spheroidal field in vacuum or spherical field in vacuum. The spheroidal parameter is chosen as $a/R=\{0,0.5,1\}$. 

For the aligned rotator, Fig.~\ref{fig:lignes_champ_B_xz_a0}, we observe the deformation of the surface as a change in the position of the foot-points of the magnetic field lines. Far from the star, especially outside the light-cylinder, there is hardly a hint about the nature of the spheroidal star. The impact is highest on the surface, and can be quantified by the polar cap rim change as will be shown in the corresponding subsection.

For the orthogonal rotator, Fig.~\ref{fig:lignes_champ_B_xy_a90}, magnetic field lines are shown in the equatorial plane. For prolate shapes, the stellar deformation is not seen because at the equator its size does not vary. The two-armed spiral pattern typical of Deutsch solution is preserved for spheroidal stars with slight changes.

As for an offset dipole or for a dipole plus multipole components, at large distances, outside the light-cylinder, the field reduces to the magnetic dipole in vacuum, washing the structure at the surface to keep only the leading lowest order term. In our case, the dominant and most relevant multipole component is the dipole, decreasing like $r^{-3}$, thus even for a spheroidal star, we expect the spin down luminosity to follow expressions very similar to a spherical star with a dependence on inclination~$\rchi$ like $\sin^2\rchi$ as will be shown in the next paragraph.

\subsection{Poynting flux}

\begin{figure}
	\centering
	\includegraphics[width=0.95\linewidth]{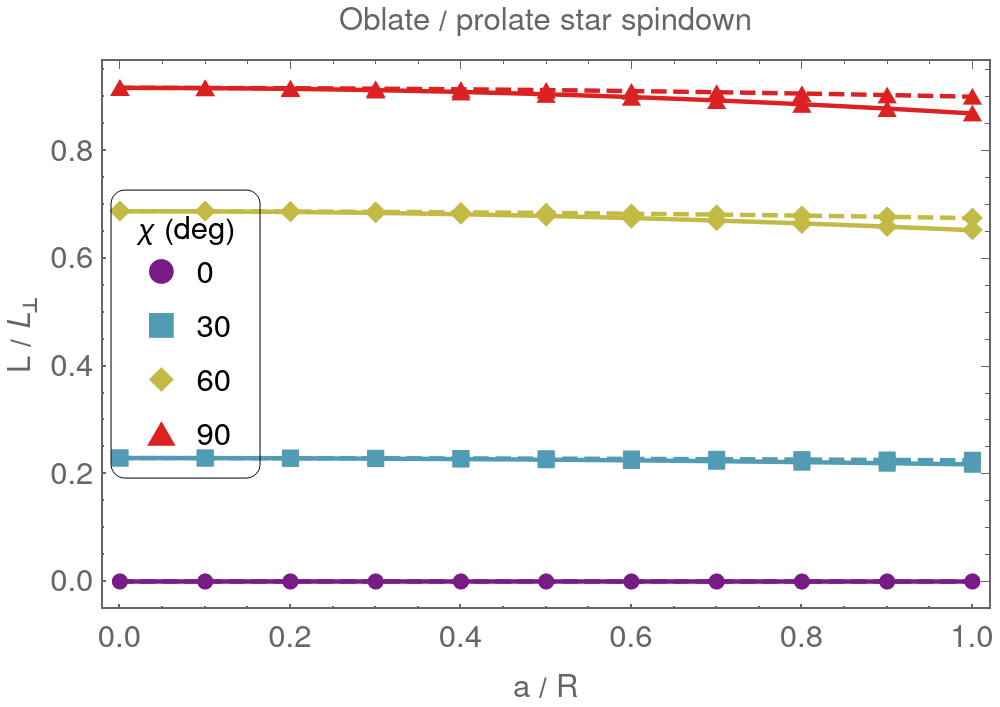}
	\caption{Spin-down luminosity for oblate and prolate stars, respectively in solid and dashed lines, with single dipole stellar boundary conditions.}
	\label{fig:luminosite_ellipticite_m112_r0.3_ro7_n129_nt32_np64_cfl0.5_ba13_alp0.1_o8_j0}
\end{figure}

\begin{figure}
	\centering
	\includegraphics[width=0.95\linewidth]{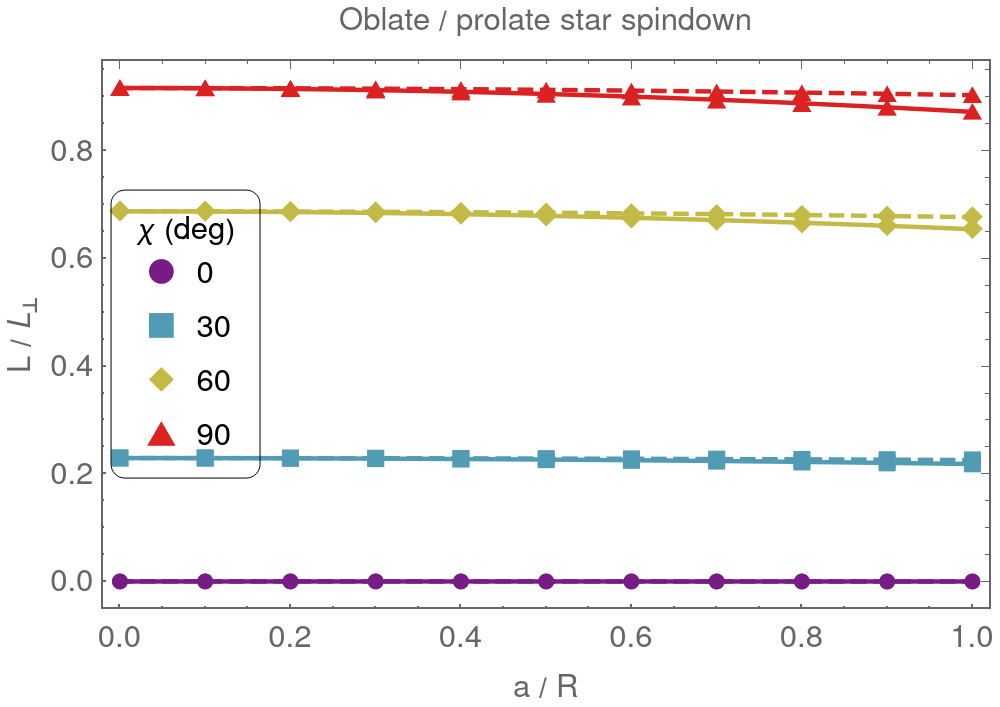}
	\caption{Spin-down luminosity for oblate and prolate stars, respectively in solid and dashed lines, with spherical dipole stellar boundary conditions.}
	\label{fig:luminosite_ellipticite_m134_r0.3_ro7_n129_nt32_np64_cfl0.5_ba13_alp0.1_o8_j0}
\end{figure}

The spin-down rate of a magnetic dipole is controlled by the Poynting flux. Exact analytical formulae exist for spherical stars but for spheroidal ones, we have to resort to numerical approximations. Fig.~\ref{fig:luminosite_ellipticite_m112_r0.3_ro7_n129_nt32_np64_cfl0.5_ba13_alp0.1_o8_j0} shows the evolution of the spin down depending on the parameter~$a$ normalized to the stellar radius~$R$ for an oblate or a prolate star, in solid and dashed lines respectively. The background magnetic field is set to the static spheroidal solution presented in Section~\ref{sec:static}. The obliquity is set to $\rchi \in \{0\degr, 30\degr, 60\degr, 90\degr\}$ as given in the legend.
Fig.~\ref{fig:luminosite_ellipticite_m134_r0.3_ro7_n129_nt32_np64_cfl0.5_ba13_alp0.1_o8_j0} shows the same evolution but for a star keeping a perfect spherical dipole structure.

As a general trend, we observe a decrease in the spin-down luminosity when the normalised asphericity~$a/R$ increases. To very good accuracy the vacuum spin-down can be approximated by quadratic corrections in $a/R$ such that
\begin{equation}
\label{eq:Fit}
 \frac{L}{L_\perp} \approx \left[ k_1 - k _2 \, \left( \frac{a}{R}\right)^2 \right] \, \sin^2\rchi .
\end{equation}
The coefficients are listed in table~\ref{tab:fit}.
\begin{table}
	\centering
\begin{tabular}{|c|c|c|}
	\hline
	Model & $k_1$ & $k_2$ \\
	\hline
	oblate & 0.921 & 0.0490  \\
	prolate & 0.921 & 0.0186  \\
	oblate spherical & 0.921 & 0.0459 \\
	prolate spherical & 0.921 & 0.0159 \\
	\hline
\end{tabular}
\caption{Fitted coefficients $k_1$ and $k_2$ as given by eq.~\eqref{eq:Fit}.\label{tab:fit}}
\end{table}
Actually the coefficient $k_1$ is known analytically and expressed in terms of spherical Hankel functions $h^{(1)}_\ell(R/\rlight)$, see \cite{petri_multipolar_2015}. For the particular values of our simulations, we should find approximately $k_1 \approx 0.919$. However, because the outgoing wave boundary conditions stand at a finite radius, relatively close to the light-cylinder, the numerical flux is impacted by these boundary surface. As carefully shown in \cite{petri_general-relativistic_2014}, the accuracy scales as $R_2^{-2}$. However, the error remains very weak, amounting to only 0.2\%.


What is the impact of this spin down on for instance stellar magnetic field inference? The accreting millisecond X-ray pulsar IGR~J00291+5934 with a spin frequency of 599~Hz is a good example, assuming a mass $M=1.4\,M_\odot$ and a radius $R=12$~km it would have $a/R \approx 0.55$ according to the MacLaurin spheroid expression in eq.~\eqref{eq:MacLaurin}. This represents a large deformation of the stellar surface. The more realistic model of \cite{silva_surface_2021} with an SLy4 equation of state would give a slightly lower value of $a/R \approx 0.48$ but still large. From equation~\eqref{eq:Lprolate} we can then find that the spheroidal formula for the luminosity gives a correction of a factor of order~2, which is significant. However from eq.~\eqref{eq:Fit} we expect a much weaker impact due to the removal of the $a/R$ dependence by our normalization procedure at large distances. This discrepancy has to be kept in mind because of the indeterminacy of a relevant normalization.

The spin down luminosity correction can be large depending on the choice of neutron star sequences used to compute the spheroidal electromagnetic field. Indeed, the normalization significantly affects the correcting factor. The key process is to choose a meaningful sequence by keeping some physical parameters constant while deforming the stellar surface from a sphere to a spheroid. Several choices are possible, for instance keeping the equatorial or the polar magnetic field strength constant as done in section~\ref{sec:static}. But we could keep the magnetic moment or the magnetic flux threading the star constant or the asymptotic field structure at large distance as done in the numerical simulations. Therefore the central question arise: to which spherical star should a spheroidal star be compared? There is no unique answer and the best normalization must be adapted to the problem under scrutiny. The spheroidal corrections to the spin down can be of the same order of magnitude as those arising from the force-free corrections which reach up to a factor~3 for the orthogonal rotator. This leads to a weaker magnetic field estimate compared to the standard vacuum magneto-dipole losses as shown by \citep{petri_illusion_2019}.


We infer that the combination of spheroidal geometry and force-free magnetosphere will lower even more the dipole magnetic field strength expectation. So how does the above fitting compare with the force-free fitting of a spherical star as found by \cite{spitkovsky_time-dependent_2006}? A quantitative answer would require computation of force-free spheroidal magnetospheres which is out of the scope of the present work. It is difficult to compare eq.~\eqref{eq:Fit} to \cite{spitkovsky_time-dependent_2006} formula because in vacuum we observe only a $L_{\rm vac} \approx L_\perp \, \sin^2 \rchi$ dependence, which has to be contrasted with the $L_{\rm FFE} \approx 3/2\,L_\perp \, (1+\sin^2 \rchi)$ dependence of the force-free model. Nevertheless we guess that both constant $\ell_1$ and $\ell_2$ in the spheroidal force-free fit $L_{\rm FFE}^{\rm spheroid} \approx 3/2\,L_\perp \, (\ell_1 + \ell_2 \, \sin^2 \rchi)$ will no longer remain constant but depend on the ratio $a/R$ at least to second order $(a/R)^2$ such that $\ell_i = \alpha_i - \beta_i \, (a/R)^2$ for $i=\{1,2\}$, $\alpha_i$ and $\beta_i$ being positive numbers with $\alpha_i \approx 1$ due to the spherical force-free results.

\subsection{Polar caps}

\begin{figure*}
	\centering
	\includegraphics[width=0.95\linewidth]{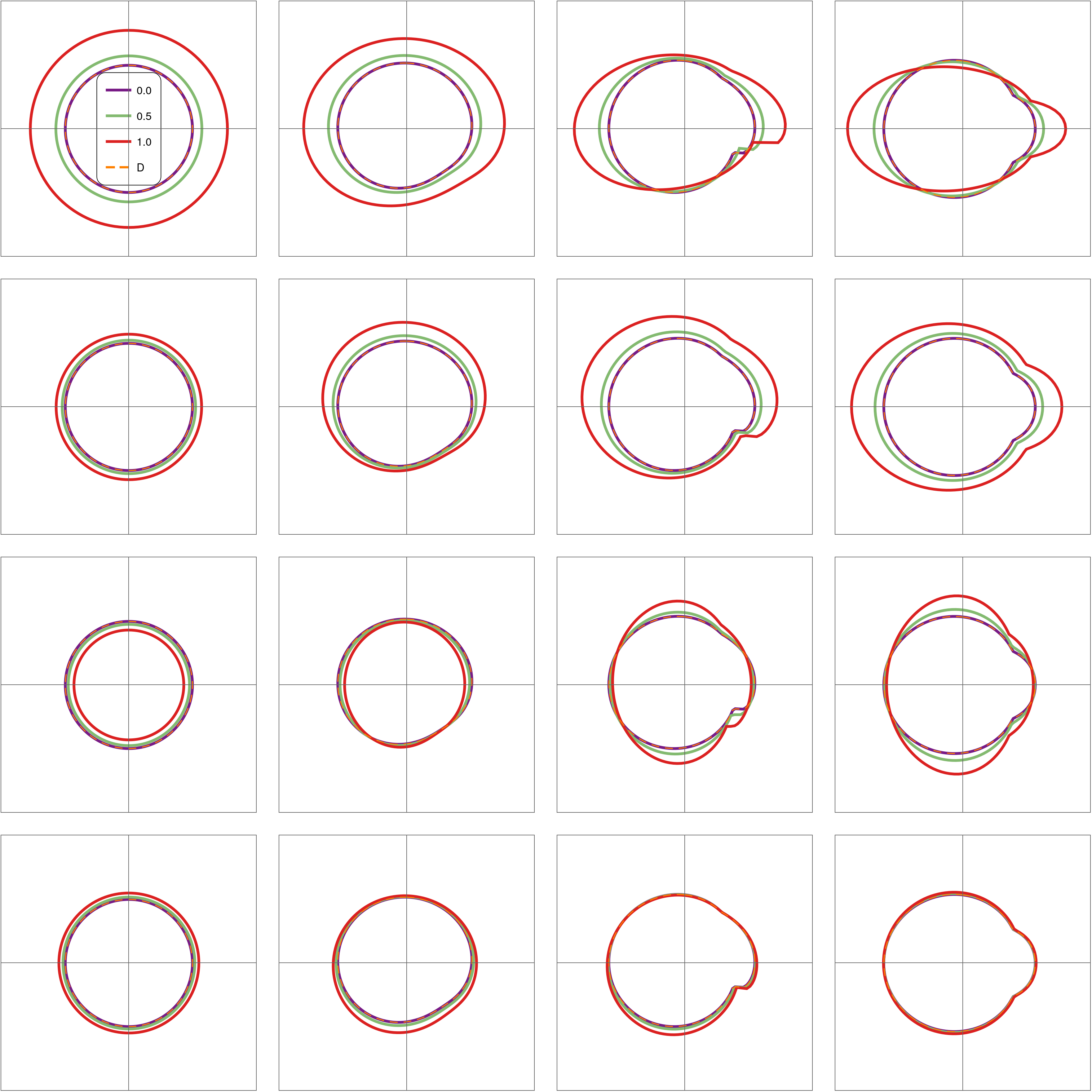}
	\caption{Polar cap shape for oblate and prolate stars with oblateness parameter $a/R=\{0,0.5,1\}$ respectively in blue, green and red. The black dashed line shows the reference solution for the Deutsch field as a check. The obliquity from the left column to the right column is $\rchi=\{0\degr, 30\degr, 60\degr, 90\degr\}$. First row for an oblate star with one mode $\ell=1$, second row for a spherical dipole magnetic field at the surface, third row for a prolate star with one mode $\ell=1$ and fourth row for a spherical dipole magnetic field at the surface.}
	\label{fig:forme_calotte}
\end{figure*}

The polar cap rims associated to the field lines structure in Fig.~\ref{fig:lignes_champ_B_xz_a0} and Fig.~\ref{fig:lignes_champ_B_xy_a90} as well as for some other obliquities are shown in Fig.~\ref{fig:forme_calotte}. We used angles $\rchi \in \{0\degr, 30\degr, 60 \degr, 90\degr\}$. As a check, the spherical case is compared to the Deutsch solution shown in orange dashed line and marked with a "D" in the legend. Both curves overlap to high precision and are indistinguishable.

The oblateness tends to elongate the polar caps in the azimuthal direction, that is in the sense of rotation and a slight contraction in the orthogonal direction for an almost perpendicular rotator. We notice an overall substantial increase in surface area for $a=R$ as seen in the first row. The second row corresponds to the same oblateness but for a spherical dipole boundary condition at the stellar surface. In this case the polar cap rim inflates in all directions whatever the obliquity.

Contrary to an oblate star, the prolate star shrinks its cap shape for almost aligned rotators as seen in the third row. While increasing the inclination angle, the polar cap elongates in the meridional direction with a slight squeezing in the sense of rotation. If spherical dipole boundary conditions are applied at the surface, the variation of polar cap shapes becomes irrelevant for an orthogonal rotator, see fourth row. This is simply explained by the fact that the stellar surface does not significantly vary around the equatorial plane for prolate shapes.

Finding realistic polar cap shapes for fast rotating neutron stars is important to model the hot spot emission seen in thermal X-rays. A careful analysis of such pulsed emission in X-ray from the NICER collaboration led to an estimate of the neutron star compactness and equatorial radius \citep{riley_nicer_2019, bogdanov_constraining_2019}. The impact of the stellar surface shape on these hot spots is discussed in \cite{silva_surface_2021}, taking also the observer orientation into account. Our simulation could help to reckon even more realistic polar caps.

\section{Conclusions}
\label{sec:conclusion}

Extending well known results from spherical magnetic stars, we showed how to express multipolar vacuum magnetic fields around spheroidal magnetized objects, being oblate or prolate. Exact analytical solutions have been derived in the case of static stars, involving Legendre functions of the third kind for the radial part, the angular part being expanded into spherical harmonics. For rotating stars, the problem cannot be solved exactly in a closed analytical form because of the introduction of radial and angular spheroidal wave functions. We showed however that some approximate solutions can be found for any realistic configuration, by an expansion into the small parameter $|\gamma|=a/\rlight \ll 1$. From a practical point of view the lowest order correction is enough to achieve high accuracy.

As a check, we also solved numerically for spheroidal rotating stars, computing the magnetic field line structure as well as the Poynting flux and the polar cap shape. This study is particular relevant for millisecond pulsars for which strong centrifugal forces inflates the equatorial part leading to an oblate shape. The change in the polar cap rim could have a significant impact on the thermal X-ray emission, modifying their light-curve in addition to general-relativistic effects like frame-dragging and light bending.

The neutron star surrounding is seldom vacuum, a pair plasma usually fills its magnetosphere, producing currents and charges modifying the electromagnetic field outside the star. We plan to add such plasma effects in the description of a spheroidal rotating magnetized celestial body.

\begin{acknowledgements}
I am grateful to the referee for helpful comments and suggestions.
This work has been supported by the CEFIPRA grant IFC/F5904-B/2018 and ANR-20-CE31-0010.
\end{acknowledgements}

\bibliographystyle{bibtex/aa}
\bibliography{/home/petri/zotero/Ma_bibliotheque}

\appendix

\section{Legendre functions of third type}
\label{app:Legendre_functions}

Legendre polynomials~$P_\ell(x)$ and their associated functions~$P_\ell^m(x)$ are usually defined in the interval~$x\in[-1,1]$. In spheroidal coordinates, we often require solutions of the Legendre differential equations outside this range. For instance when studying fields outside an object up to infinity we require $x>1$. We are particularly interested in solutions decreasing to zero at large distances, meaning an electric and magnetic potential falling to zero when $x\to +\infty$. These solutions are then called Legendre functions of the third type~$Q_\ell^m(x)$ and related to prolate coordinates for $|x|\leq1$ by analytical continuation in the complex plane, starting from $x\in[-1,1]$. They are solutions of the second order linear differential equation
\begin{equation}\label{eq:EDO1}
\frac{d}{dx} \left[(1-x^2)\, \frac{dQ_\ell^m(x)}{dx}\right] + \left( \ell \, (\ell+1) - \frac{m^2}{1-x^2}\right) \, Q_\ell^m(x) = 0
\end{equation}
where $\ell$ and $m$ are two integers. It corresponds to the radial part of a multipole of order $(\ell,m)$ in prolate spheroidal coordinates and related to the spherical harmonic $Y_\ell^m(\theta,\phi)$.

For practical purposes, we list the first few useful functions for the monopole $\ell=0$, the dipole $\ell=1$ and the quadrupole $\ell=2$, which are real-valued and given by
\begin{subequations}
	\begin{align}
		{Q}_0^0(x) & = \textrm{arccoth } x \\
		{Q}_1^0(x) & = x \, \textrm{arccoth } x -1 \\
		{Q}_1^1(x) & = \frac{(x^2-1)\,\textrm{arccoth } x -x}{\sqrt{x^2-1}} \\
		{Q}_2^0(x) & = \frac{(3\,x^2-1) \, \textrm{arccoth } x - 3 \, x}{2} \\
		{Q}_2^1(x) & = \frac{3\,x\,(x^2-1)\,\textrm{arccoth } x + 2 - 3\,x^2}{\sqrt{x^2-1}} \\
		{Q}_2^2(x) & = 3\,(x^2-1)\,\textrm{arccoth } x + x \, \frac{5-3\,x^2}{x^2-1} . 
	\end{align}
\end{subequations}
These expressions are used to compute the radial profile of the electric or magnetic potential in prolate spheroidal coordinates.

For oblate spheroidal coordinates, we require solutions $\mathrm{Q}_\ell^m$ for $x>1$ such that
\begin{equation}\label{eq:EDO2}
\frac{d}{dx} \left[(1+x^2)\, \frac{d\mathrm{Q}_\ell^m(x)}{dx}\right] + \left( \frac{m^2}{1+x^2} - \ell \, (\ell+1) \right) \, \mathrm{Q}_\ell^m(x) = 0
\end{equation}
with the correspondence $\mathrm{Q}_\ell^m(x) = i^\ell \, Q_\ell^m(i\,x)$. Note the change in sign in front of the factor $x^2$ of eq.~\eqref{eq:EDO2} compared to  eq.~\eqref{eq:EDO1}.

For practical purposes, here also we list the first few useful functions for the monopole $\ell=0$, the dipole $\ell=1$ and the quadrupole $\ell=2$, which are again all real-valued and given by
\begin{subequations}
	\begin{align}
	\mathrm{Q}_0^0(x) & = \arccot x \\
	\mathrm{Q}_1^0(x) & = 1 - x \, \arccot x \\
	\mathrm{Q}_1^1(x) & = \frac{x-(1+x^2)\,\arccot x}{\sqrt{1+x^2}} \\
	\mathrm{Q}_2^0(x) & = \frac{(1+3\,x^2) \, \arccot x - 3\,x}{2}  \\
	\mathrm{Q}_2^1(x) & = \frac{3\,x\,(1+x^2)\,\arccot x - 2 - 3\,x^2}{\sqrt{1+x^2}} \\
	\mathrm{Q}_2^2(x) & = 3\,(1+x^2)\,\arccot x - x \, \left( 3 + \frac{2}{1+x^2}\right) .
	\end{align}
\end{subequations}
These expressions are used to compute the radial profile of the electric or magnetic potential in oblate spheroidal coordinates.

\section{Angular functions expansion}
\label{app:Angular_functions_expansion}

The prolate angular wave functions~$\mathrm{Ps}_\ell^m(\gamma,\eta)$ are expanded into Legendre functions $P_\ell^m(\eta)$ according to
\begin{equation}
\mathrm{Ps}_\ell^m(\gamma,\eta) = {\sum_{r=0,1}^{+\infty}}' d_r^{m\ell} \, P_{m+r}^m(\eta) .
\end{equation}
where the prime indicates summation from 0/1 over even/odd indices when $(\ell-m)$ is even/odd.
The expansion coefficients $d_r^{m\ell}$ are determined by solving a three-term recurrence relation with coefficients given in \cite{abramowitz_handbook_1965}. The lowest order corrections in $\gamma^2$ are
\begin{subequations}
	\begin{align}
	d_{\ell-m}^{m\ell} & \approx 1 + O(\gamma^4) \\
	d_{\ell-m-2}^{m\ell} & \approx - \frac{(\ell + m - 1) \, (\ell + m)}{2\,(2\, n-1)^2 \, (2\, n +1)} \, \gamma^2  + O(\gamma^4) \\
	d_{\ell-m+2}^{m\ell} & \approx - \frac{(\ell - m + 1) \, (\ell - m + 2)}{2\,(2\, n+3)^2 \, (2\, n +1)} \, \gamma^2  + O(\gamma^4).
	\end{align}
\end{subequations}
If the subscript is negative, the coefficient vanishes by convention.
To this level of approximation, we neglect corrections being of order $\gamma^4$ or higher and therefore no corrections apply to the dominant coefficient~$d_{\ell-m}^{m\ell}\approx1$.

For the expansion of the first multipoles with $m=1$, we give the expression of $d_{r}^{m\ell}$ to $\gamma^2$ order for $\ell-m=\pm2$ for $m=1$.
\begin{subequations}
	\begin{align}
	& d_{2}^{1,1}  \approx \frac{\gamma^2}{75} + O(\gamma^4) \\
	& d_{3}^{1,2}  \approx \frac{3\,\gamma^2}{245} + O(\gamma^4) \\
	d_{0}^{1,3} \approx -\frac{6\,\gamma^2}{175} + O(\gamma^4) \qquad ; \qquad & d_{4}^{1,3}  \approx \frac{2\,\gamma^2}{189} + O(\gamma^4)   \\
	d_{1}^{1,4} \approx -\frac{10\,\gamma^2}{441} + O(\gamma^4)\qquad ; \qquad& d_{5}^{1,4}  \approx \frac{10\,\gamma^2}{1089} + O(\gamma^4)   .
	\end{align}
\end{subequations}
Explicitly, for the first wave functions, this means
\begin{subequations}
	\begin{align}
	\mathrm{Ps}_1^1(\gamma,\eta) & \approx P_1^1(\eta) + \frac{\gamma^2}{75} \, P_3^1(\eta) + O(\gamma^4) \\
	\mathrm{Ps}_2^1(\gamma,\eta) & \approx P_2^1(\eta) + \frac{3\,\gamma^2}{245} \, P_4^1(\eta) + O(\gamma^4) \\
	\mathrm{Ps}_3^1(\gamma,\eta) & \approx P_3^1(\eta) - \frac{6\,\gamma^2}{175} \, P_1^1(\eta) + \frac{2\,\gamma^2}{189} \, P_5^1(\eta)  + O(\gamma^4) \\
	\mathrm{Ps}_4^1(\gamma,\eta) & \approx P_4^1(\eta) - \frac{10\,\gamma^2}{441} \, P_2^1(\eta) +  \frac{10\,\gamma^2}{1089} \, P_6^1(\eta) + O(\gamma^4) 
	\end{align}
\end{subequations}

\end{document}